\documentclass[preprint,showpacs,preprintnumbers]{revtex4}
\usepackage{amssymb}
\usepackage{graphicx}
\usepackage{dcolumn}
\usepackage{bm}

\begin{document}

\title{Quantum Potential for Diffraction and Exchange Effects}
\author{James Dufty and Sandipan Dutta }
\affiliation{Department of Physics, University of Florida, Gainesville, FL 32611}
\author{Michael Bonitz and Alexei Filinov}
\affiliation{Institut f\"ur Theoretische Physik und Astrophysik,
Christian-Albrechts-Universit\"at zu Kiel, D-24098 Kiel, Germany}
\date{\today }

\begin{abstract}
Semi-classical methods of statistical mechanics can incorporate essential
quantum effects by using effective quantum potentials. An ideal Fermi gas
interacting with an impurity is represented by a classical fluid with
effective electron-electron and electron-impurity quantum potentials. The
electron-impurity quantum potential is evaluated at weak coupling, leading
to a generalization of the Kelbg potential to include both diffraction and
degeneracy effects. The electron-electron quantum potential for exchange
effects only is the same as that discussed earlier by others.
\end{abstract}

\maketitle

\section{Introduction}

The application of classical Newtonian mechanics to materials is generally
limited to conditions of small characteristic quantum wavelengths (e.g.,
high temperatures, large mass). In some cases (e.g., electron-proton
systems) there is no simple classical limit due to the singular attractive
interaction. Still, it is useful to explore possible realizations of an
inherently quantum description as a semi-classical problem. This cannot be
done in general, but exact or approximate correspondences can be made for
specific properties. The advantage of such semi-classical realizations is
that powerful classical methods can be employed to address the difficult
many-body problem (e.g., Monte Carlo integration methods for partition
functions represented in terms of classical actions, molecular dynamics (MD)
implementation of Newton's equations).

One approach that has met with significant success is to replace the given
interaction potential with an effective "quantum potential" in a
corresponding classical description. The quantum potential incorporates some
or all of the important quantum effects in a modification of its functional
form. In the case of pairwise additive potentials, a quantum potential has
been defined for equilibrium calculations in terms of the exact two particle
density matrix for a given pair of particles by equating it to the
corresponding classical form with an effective potential. In this way, the
quantum potential incorporates the quantum diffraction effects and other
non-classical features such as binding energies. A practical form is
obtained by a first order expansion of the quantum potential in terms of the
given potential, leading to the Kelbg potential \cite{kelbg}. In the case of
the Coulomb interaction, the Kelbg form shows a "regularization" of the
short range singularity by a smoothing of the potential over distances of
the order of the thermal de Broglie wavelength. Important applications of
these potentials include MD simulations for a Hydrogen plasma, and
construction of an action for the singular Coulomb interactions to allow
path integral Monte Carlo (PIMC) \cite{PIMC} evaluation of quantum partition
functions. More general non-perturbative methods to determine such a quantum
potential from the two particle density matrix have been explored and tested
\cite{feynman,filinov}. Extensions of these ideas to external forces \cite%
{external}\ and non-equilibrium states also have been discussed \cite{fromm}.

The most important cases of interest involve electrons under conditions
where quantum degeneracy can be important. Quantum potentials based on the
two particle density matrix do not account for many-electron exchange
effects. An important early study of this problem was the construction by
Lado of a classical system incorporating the exchange effects of an ideal
quantum gas \cite{Lado}. The classical gas has pairwise additive quantum
potentials chosen to give the correct quantum electron - electron pair
correlation function. This was accomplished using the second equation of the
Born-Green hierarchy \cite{Hansen}, solved for the quantum potential in
terms of the known correlation functions. This idea has been given a more
practical form with the role of the second Born-Green equation replaced by
the hypernetted chain (HNC) integral equation approximation \cite{Hansen}
relating the correlation function to the quantum potential \cite{Perrot}.
Subsequetly, the interacting quantum system is represented by an extended
quantum potential that is the sum of that described above for exchange
effects plus a regularized real potential of interaction with diffraction
effects. The objective here is to illustrate the simplest case in which the
effects of degeneracy and diffraction appear coupled, rather than additive.
The system considered is again the ideal Fermi gas, but with the addition of
an impurity interacting with each particle. The corresponding classical
system has an electron-electron quantum potential as described by Lado for
exchange, and an additional electron-impurity interaction with both exchange
and diffraction effects. An additional Born-Green equation for the
electron-impurity quantum potential entails a new correlation function for
the impurity with both diffraction and exchange effects, as well as coupling
to the electron-electron quantum potential. This equation is solved for weak
coupling conditions, leading to the Kelbg result in the non-degenerate limit
but more generally describing coupled exchange and diffraction effects. For
the case of Coulomb coupling to the impurity, it is shown that the
degeneracy effects can be described to good approximation by an appropriate
scaling of the Kelbg functional form.

There are many different ways in which attempts have been made to introduce
quantum effects into classical descriptions, so it is important to clarify
the context of the present calculations. First, they are among a class of
quantum potentials that are based on equilibrium properties and pairwise
additivity. Their use in molecular dynamics simulations for nonequilibrium
states and for transport properties are therefore uncontrolled. Three-body
and many-body quantum effects are not included so the formation of bound
pairs may be described accurately \cite{filinov} but more complex molecular
structures are outside the realm of accuracy. Representations involving
many-body quantum potentials follow directly from truncated cluster
expansions of the Slater sum and exact field theoretical representations
such as a classical polymer action come at the price of considerable
additional complexity. Quantum potentials not tied to the equilibrium state,
such as those from wave-packet molecular dynamics have a potentially wider
domain of applicability, but also entail a new level of phenomenology. A
more controlled introduction of momentum dependent quantum forces from the
Wigner representation of the von Neumann equation are specific to each
state, equililbrium or non-equilibrium, but are still in an early state of
exploration. A closely related field is that of quantum hydrodynamics. Some
of the diversity of issues around quantum potentials have been critiqued
recently \cite{Jones}

It is a pleasure to dedicate this contribution to Frank Harris - exceptional
mentor, colleague, and friend to all fortunate enough to have crossed paths
with him.

\section{Quantum potentials for impurity in an ideal Fermi gas}

Consider a system of $N$ non-interacting electrons at equilibrium in an
impurity field fixed (e.g., infinite mass) at the origin. The Hamiltonian
operator is%
\begin{equation}
\widehat{H}=\sum_{\alpha =1}^{N}\left( \frac{\widehat{p}_{\alpha }^{2}}{2m}%
+V(\widehat{q}_{\alpha })\right) ,  \label{2.1}
\end{equation}%
where $V(\widehat{q}_{\alpha })$ is the central potential due to the
impurity at the position $\widehat{\mathbf{q}}_{\alpha }$ of electron $%
\alpha $. A caret over a symbol is used to distinguish an operator from its
corresponding classical variable. The average number density at a distance $%
\mathbf{r}$ from the impurity in the Grand Canonical ensemble is%
\begin{equation}
n_{ei}(r;z,\beta )=<\widehat{n}(\mathbf{r})>=\frac{1}{\mathcal{Z}}%
\sum_{N}z^{N}Tre^{-\beta \widehat{H}}\widehat{n}(\mathbf{r}).  \label{2.2}
\end{equation}%
Here $Tr$ denotes a trace over a complete set of anti-symmetrized $N$
electron states. Also, the partition function $\mathcal{Z}$ and number
operator $\widehat{n}$ are
\begin{equation}
\mathcal{Z}(z,\beta )=\sum_{N}z^{N}Tre^{-\beta \widehat{H}},\hspace{0.25in}%
\widehat{n}(\mathbf{r})=\sum_{\alpha =1}^{N}\delta \left( \mathbf{r}-%
\widehat{\mathbf{q}}_{\alpha }\right) ,  \label{2.3}
\end{equation}%
$\beta =1/k_{B}T$ is the inverse temperature, and $z$ is related to the
chemical potential $\mu $ by $z=e^{\beta \mu }$. Similarly, the pair density
for two electrons at distances $\mathbf{r}$ and $\mathbf{r}^{\prime }$ from
the impurity is given by
\begin{equation}
n_{eei}(\mathbf{r},\mathbf{r}^{\prime };z,\beta )=<\left( \widehat{n}(%
\mathbf{r})\widehat{n}(\mathbf{r}^{\prime })-\delta \left( \mathbf{r}-%
\mathbf{r}^{\prime }\right) \widehat{n}(\mathbf{r})\right) >.  \label{2.4}
\end{equation}%
Finally, all correlation functions for electron densities at arbitrary
positions not referred to the location of the impurity become independent of
the impurity in the thermodynamic limit and therefore are just those for the
ideal Fermi gas, e.g.%
\begin{equation}
n_{ee}(\left\vert \mathbf{r}-\mathbf{r}^{\prime }\right\vert ;z,\beta
)=n_{eei}(\mathbf{r},\mathbf{r}^{\prime };z,\beta )\mid _{V=0.}  \label{2.5}
\end{equation}

A corresponding representative classical system is defined by the Hamiltonian

\begin{equation}
H_{cl}=\sum_{\alpha =1}^{N}\left( \frac{p_{\alpha }^{2}}{2m}+\mathcal{V}%
_{ei}(q_{\alpha })\right) +\frac{1}{2}\sum_{\alpha ,\sigma =1}^{N}\mathcal{V}%
_{ee}(\left\vert \mathbf{q}_{\alpha }-\mathbf{q}_{\sigma }\right\vert ).
\label{2.6}
\end{equation}%
The "quantum" potentials $\mathcal{V}_{ei}$ and $\mathcal{V}_{ee}$ are
chosen to assure that the classical system preserves key properties of the
underlying quantum system. A natural choice is the requirement that the
classical electron density about the impurity $n_{ei}(\mathbf{r};z,\beta )$
and the classical electron-electron pair density $n_{ee}(\mathbf{r},\mathbf{r%
}^{\prime };z,\beta )$ be the same as those for the quantum system. This
requires calculation of the classical expressions for $n_{ei}$ and $n_{ee}$
for the Hamiltonian (\ref{2.6}) as functionals of the quantum potentials,
equating these expressions to the corresponding quantum expressions, and
inverting those equalities to find $\mathcal{V}_{ei}$ and $\mathcal{V}_{ee}$
as functionals of the quantum $n_{ei}$ and $n_{ee}$. Although calculation of
the quantum expressions is straightforward (but non-trivial for $n_{ei}$),
the corresponding classical calculation confronts the full many-body problem
due to the pair interactions in (\ref{2.6}). Lado approached this problem by
considering the exact Born-Green equations obeyed by the classical forms for
$n_{ei}(\mathbf{r};z,\beta )$ and $n_{ee}(\mathbf{r},\mathbf{r}^{\prime
};z,\beta )$

\begin{equation}
\mathbf{\nabla }_{1}n_{ei}\left( r_{1}\right) =-\beta n_{ei}\left(
r_{1}\right) \mathbf{\nabla }_{1}\mathcal{V}_{ei}\left( r_{1}\right) -\beta
\int d\mathbf{r}_{2}n_{eei}\left( \mathbf{r}_{1},\mathbf{r}_{2}\right)
\mathbf{\nabla }_{1}\mathcal{V}_{ee}\left( r_{21}\right) ,  \label{2.7}
\end{equation}%
\begin{equation}
\mathbf{\nabla }_{1}n_{ee}\left( r_{12}\right) =-\beta n_{ee}\left(
r_{12}\right) \mathbf{\nabla }_{1}\mathcal{V}_{ee}\left( r_{12}\right)
-\beta \int d\mathbf{r}_{3}n_{eee}\left( \mathbf{r}_{1},\mathbf{r}_{2},%
\mathbf{r}_{3}\right) \mathbf{\nabla }_{1}\mathcal{V}_{ee}\left(
r_{31}\right) .  \label{2.8}
\end{equation}%
These equations are part of an infinite hierarchy, coupling correlations
among $m$ particles to those for $m+1$. For example, (\ref{2.8}) relates $%
n_{ee}\left( r_{12}\right) $ to the quantum potential $\mathcal{V}%
_{ee}\left( r_{12}\right) $, as desired, but also couples it to $%
n_{eee}\left( \mathbf{r}_{1},\mathbf{r}_{2},\mathbf{r}_{3}\right) $. In the
present context, $n_{ee}\left( r_{12}\right) $ is replaced by the known
quantum form, but $n_{eee}\left( \mathbf{r}_{1},\mathbf{r}_{2},\mathbf{r}%
_{3}\right) $ must still be calculated as a functional of the quantum
potential. Then (\ref{2.8}) can be solved for $\mathcal{V}_{ee}\left(
r_{12}\right) $. Thus, the difficult many-body problem reappears in the need
to calculate $n_{eee}\left( \mathbf{r}_{1},\mathbf{r}_{2},\mathbf{r}%
_{3}\right) $. A similar difficulty is clearly present in equation (\ref{2.7}%
) for $n_{ei}\left( r_{1}\right) $.

Lado avoided the classical determination of $n_{eee}$ by using the
corresponding quantum correlation function, a much easier ideal gas
calculation \cite{Lado}. Then (\ref{2.8}) becomes a simple linear integral
equation that can be solved for $\mathcal{V}_{ee}$ numerically. However,
this use of the quantum expression for $n_{eee}$ introduces a new
approximation since (\ref{2.8}) follows from the classical Hamiltonian in
terms of the classical form for $n_{eee}$ as a functional of $\mathcal{V}%
_{ee}$. There is no reason to expect that the classical and quantum forms
should be the same. An alternative approach \cite{Perrot} has been suggested
more recently based on a classical "closure" expressing $n_{eee}$ in terms
of $n_{ee}\left( r_{12}\right) $ and $\mathcal{V}_{ee}$, the hypernetted
chain (HNC) approximation \cite{Hansen}. This is an approximation to the
classical many-body problem and therefore more self-consistent than the Lado
approach. In practice, it is found that results obtained by both methods are
quite close.

Since (\ref{2.8}) is determined independently of the impurity it will not be
considered further here, and $\mathcal{V}_{ee}$ will be considered as known
for the purposes of solving (\ref{2.7}). The latter has similar problems to
that just described, namely determination of the classical form for $n_{eei}$%
. In addition, the quantum form for $n_{ei}$ is more difficult, requiring
construction from the eigenvalues and eigenfunctions for an electron in the
presence of the ion. This is similar to the problem considered by Kelbg for
the two particle density matrix. He simplified the problem by considering
weak coupling conditions, and the same will be done here in the remainder of
the manuscript. Weak coupling here means $\beta V<<1$ so that functional
expansion of $\mathcal{V}_{ee}$, $n_{ei}\left( r_{1}\right) ,$ and $%
n_{eei}\left( \mathbf{r}_{1},\mathbf{r}_{2}\right) $ can be exploited. This
is described in the next subsection.

\subsection{Weak coupling}

It can be shown from (\ref{2.7}) that $\mathcal{V}_{ei}$ vanishes if $V=0$,
and so can be written
\begin{equation}
\beta \mathcal{V}_{ei}(r\mid V)=\int d\mathbf{r}^{\prime }G(\left\vert
\mathbf{r-r}^{\prime }\right\vert )\beta V(r^{\prime })+..  \label{2.9}
\end{equation}%
The dots denote second and higher orders in $\beta V$. Similarly,%
\begin{equation}
n_{ei}\left( r\right) =n_{e}\left( z,\beta \right) +\int d\mathbf{r}^{\prime
}\frac{\delta n_{ei}\left( r\right) }{\delta V(r^{\prime })}\mid
_{V=0}V(r^{\prime })+..  \label{2.10}
\end{equation}%
where $n_{e}\left( z,\beta \right) $ is the ideal Fermi gas density.
Finally, the classical definition for $n_{eei}\left( \mathbf{r}_{1},\mathbf{r%
}_{2}\right) $ for the Hamiltonian (\ref{2.6}) gives the corresponding
expansion
\begin{eqnarray}
n_{eei}\left( \mathbf{r}_{1},\mathbf{r}_{2}\right) &=&n_{ee}\left(
r_{12}\right) \left( 1-\beta \mathcal{V}_{ei}\left( r_{1}\right) -\beta
\mathcal{V}_{ei}\left( r_{2}\right) \right)  \nonumber \\
&&-\int d\mathbf{r}_{3}\beta \mathcal{V}_{ei}\left( r_{3}\right)
n_{eee}\left( \mathbf{r}_{1},\mathbf{r}_{2},\mathbf{r}_{3}\right) +..
\label{2.11}
\end{eqnarray}%
Substitution of (\ref{2.11}) into the second term on the right side of (\ref%
{2.7}), and use of (\ref{2.8}) gives the simplification
\begin{equation}
\int d\mathbf{r}_{2}n_{eei}\left( \mathbf{r}_{1},\mathbf{r}_{2}\right)
\mathbf{\nabla }_{1}\mathcal{V}_{ee}\left( r_{21}\right) \mathbf{=}\int d%
\mathbf{r}_{2}\mathcal{V}_{ei}\left( r_{2}\right) \mathbf{\nabla }%
_{1}n_{ee}\left( r_{12}\right) +..  \label{2.12}
\end{equation}%
With these results, (\ref{2.7}) can be expanded to first order in $V$ giving
the desired expression for the function $G(\left\vert \mathbf{r-r}^{\prime
}\right\vert )$ that determines $\mathcal{V}_{ei}$ in (\ref{2.9}) to leading
order
\begin{equation}
-\beta ^{-1}\frac{\delta n_{ei}\left( r_{1}\right) }{\delta V(\mathbf{r}_{2})%
}\mid _{V=0}=n_{e}G(\mathbf{r}_{1}\mathbf{-r}_{2})+\int d\mathbf{r}%
_{3}\left( n_{ee}\left( r_{13}\right) -n_{e}^{2}\right) G(\mathbf{r}_{3}%
\mathbf{-r}_{2})  \label{2.13}
\end{equation}%
where the ideal gas functions $n_{e}$ and $n_{ee}\left( r\right) $ are
\begin{equation}
n_{e}=\frac{\left( 2s+1\right) }{h^{3}}\int d\mathbf{p}n(p),\hspace{0.25in}%
n(p)=\left( z^{-1}e^{\beta p^{2}/2m}+1\right) ^{-1},  \label{2.14}
\end{equation}%
\begin{equation}
n_{ee}\left( r\right) =n_{e}^{2}-\left( 2s+1\right) \left( \frac{1}{h^{3}}%
\int d\mathbf{p}n(p)e^{\frac{i}{\text{%
h{\hskip-.2em}\llap{\protect\rule[1.1ex]{.325em}{.1ex}}{\hskip.2em}%
}}\mathbf{p\cdot r}}\right) ^{2}.  \label{2.15}
\end{equation}%
Here, $s$ is the spin of the Fermions.

The response function $\beta ^{-1}\delta n_{ei}\left( r_{1}\right) /\delta V(%
\mathbf{r}_{2})$ on the left side of (\ref{2.13}) describes the direct
effects of exchange and diffraction on the electron interacting with the
impurity. In addition, this couples via the second term on the right to the
exchange effects among electrons not interacting with the impurity (i.e. a
coupling of $\mathcal{V}_{ei}$ to $\mathcal{V}_{ee}$ in (\ref{2.7})). This
coupling is essential to describe the degeneracy of the background ideal
quantum gas. To illustrate this, note that for the special case of $V(r)$
constant, $\mathcal{V}_{ei}(r\mid V)\rightarrow V$ since in that case $V$
simply gives a shift of the chemical potential. Therefore, in general%
\begin{equation}
\int d\mathbf{r}G(r)=1.  \label{2.16}
\end{equation}%
Integrating (\ref{2.13}) then gives%
\begin{eqnarray}
-\beta ^{-1}\int d\mathbf{r}^{\prime }\frac{\delta n(\mathbf{r},z,\beta \mid
V)}{\delta V(\mathbf{r}^{\prime })} &\mid &_{V=0}=\frac{\partial n_{e}}{%
\partial \ln z}=n_{e}+\int d\mathbf{r}_{3}\left( n_{ee}\left( r_{13}\right)
-n_{e}^{2}\right)  \nonumber \\
&=&n_{e}-\frac{\left( 2s+1\right) }{h^{3}}\int d\mathbf{p}n^{2}(p).
\label{2.17}
\end{eqnarray}%
The second line follows from the definitions (\ref{2.14}) and (\ref{2.15}),
confirming that the right side is indeed the derivative on the left. Thus,
it is seen that the coupling of $\mathcal{V}_{ei}$ to $\mathcal{V}_{ee}$ is
essential for consistency with the quantum thermodynamics.

It is now straightforward to calculate the response function $\beta
^{-1}\delta n_{ei}\left( r_{1}\right) /\delta V(\mathbf{r}_{2})$ at $V=0$
and to solve (\ref{2.13}) for $G$ by Fourier transformation. The
corresponding Fourier transformed potential $\widetilde{\mathcal{V}}_{ei}(k)$
from (\ref{2.9}) is found to be
\begin{equation}
\widetilde{\mathcal{V}}_{ei}(k)=\widetilde{G}(k)\widetilde{V}(k)+..,\hspace{%
0.25in}\widetilde{G}(k)=\frac{\text{Re }\Pi _{0}(k,\omega =0)}{\text{Re }\Pi
_{0}(0,\omega =0)}\frac{\left( 1+d_{ee}\left( 0\right) \right) }{\left(
1+d_{ee}\left( k\right) \right) }  \label{2.18}
\end{equation}%
where $\Pi _{0}(k,\omega )$ is the polarization function for the ideal Fermi
gas from finite temperature Greens function theory \cite{fetter}
\begin{equation}
\widetilde{\Pi }_{0}(k,\omega )=\lim_{\eta \rightarrow 0}2\int \frac{d%
\mathbf{p}}{\left( 2\pi \right) ^{3}}\frac{n(p)-n\left( \left\vert \mathbf{p}%
-\text{%
h{\hskip-.2em}\llap{\protect\rule[1.1ex]{.325em}{.1ex}}{\hskip.2em}%
}\mathbf{k}\right\vert \right) }{\omega +i\eta +e\left( p\right) -e\left(
\left\vert \mathbf{p}-\text{%
h{\hskip-.2em}\llap{\protect\rule[1.1ex]{.325em}{.1ex}}{\hskip.2em}%
}\mathbf{k}\right\vert \right) },  \label{2.19}
\end{equation}%
\begin{equation}
e\left( p\right) =\frac{p^{2}}{2m},\hspace{0.25in}n(p)=\left( z^{-1}e^{\beta
e\left( p\right) }+1\right) ^{-1},  \label{2.20}
\end{equation}%
and $d_{ee}\left( k\right) $ represents the effects of coupling to $\mathcal{%
V}_{ee}$
\begin{equation}
d_{ee}\left( k\right) =\frac{1}{n_{e}}\int d\mathbf{r}e^{i\mathbf{k\cdot r}%
}\left( n_{ee}\left( r_{13}\right) -n_{e}^{2}\right) =-\frac{\left(
2s+1\right) }{n_{e}h^{3}}\int d\mathbf{p}n(p)n(\left\vert \mathbf{p}-\text{%
h{\hskip-.2em}\llap{\protect\rule[1.1ex]{.325em}{.1ex}}{\hskip.2em}%
}\mathbf{k}\right\vert ).  \label{2.21}
\end{equation}%
Note that $1+d_{ee}\left( k\right) =S_{ee}(k)$ is the ideal Fermi gas static
structure factor. The quantum potential given by (\ref{2.18}) is quite
general, applying at weak coupling but for arbitrary degree of degeneracy.

\section{Coulomb interaction}

An important special case is the Coulomb potential (e.g. a point ion at the
origin), $V(r)=qe/r,$ where $e$ is the magnitude of the electron charge and
the impurity charge $q$ can be negative or positive. In the following the $k$
dependence of the coupling to $\mathcal{V}_{ee}$ in (\ref{2.21}) will be
neglected (but not its coupling for $k=0$). Then the inverse transform of (%
\ref{2.18}) can be performed exactly \cite{dufty} to determine $\mathcal{V}%
_{ei}(r)$ with the result%
\begin{equation}
\mathcal{V}_{ei}(r)\rightarrow V(r)S(\frac{r}{\lambda },z)  \label{3.16}
\end{equation}%
where $\lambda =\sqrt{2\pi \text{%
h{\hskip-.2em}\llap{\protect\rule[1.1ex]{.325em}{.1ex}}{\hskip.2em}%
}^{2}\beta /m}$ is the thermal de Broglie wavelength and the quantum
regularization effect $S(r/\lambda ,z)$ is
\begin{equation}
S(\frac{r}{\lambda },z)=\frac{r}{\lambda }\int_{0}^{\infty }dxn^{\ast
}(x,z)\left( \frac{\lambda }{r}\left( 1-\cos \left( \frac{4x\sqrt{\pi }r}{%
\lambda }\right) \right) +4x\sqrt{\pi }\left( \frac{1}{2}\pi -\text{{Si}}%
\left( \frac{4x\sqrt{\pi }r}{\lambda }\right) \right) \right) .  \label{3.17}
\end{equation}%
Also $n^{\ast }(x,z)$ is the dimensionless Fermi function normalized to
unity and Si$\left( x\right) $ is the sine integral
\begin{equation}
n^{\ast }(x,z)\equiv \frac{1}{z^{-1}e^{x^{2}}+1}\left( \int_{0}^{\infty }dx%
\frac{1}{\left( 1+z^{-1}e^{x^{2}}\right) }\right) ^{-1},\hspace{0.25in}\text{%
Si}\left( x\right) =\int_{0}^{x}dx^{\prime }\frac{\sin x^{\prime }}{%
x^{\prime }}  \label{3.18}
\end{equation}

It is easily verified that $S(r/\lambda ,z)$ is proportional to $r$ for
small $r/\lambda $%
\begin{equation}
S(r/\lambda ,z)\rightarrow 2\pi \frac{r}{\lambda (z)},\hspace{0.25in}\lambda
\left( z\right) \equiv \frac{\lambda }{\sqrt{\pi }\int_{0}^{\infty
}dxxn^{\ast }(x,z)},  \label{3.19}
\end{equation}%
so that the Coulomb divergence is removed. Also $S(r/\lambda ,z)\rightarrow
1 $ for large $r$ so that the Coulomb potential is recovered, as required by
(\ref{2.16}). Finally in the non-degenerate limit, $z\rightarrow 0$, and the
Kelbg result is obtained \cite{kelbg}

\begin{equation}
S(\frac{r}{\lambda },z)\rightarrow S_{K}(\frac{r}{\lambda })\equiv
1-e^{-4\pi \left( r/\lambda \right) ^{2}}-2\pi \frac{r}{\lambda }\left(
\text{erf}\left( \frac{2\sqrt{\pi }r}{\lambda }\right) -1\right) .
\label{3.20}
\end{equation}%
In the opposite limit of strong degeneracy, $z>>1$, (\ref{3.17}) gives%
\begin{eqnarray}
S(\frac{r}{\lambda },z) &\rightarrow &1-\frac{\lambda }{4\sqrt{\pi \ln z}r}%
\sin \left( \frac{4\sqrt{\pi \ln z}r}{\lambda }\right)  \nonumber \\
&&+\frac{1}{2}\frac{4\sqrt{\pi \ln z}r}{\lambda }\left( j_{1}\left( \frac{4%
\sqrt{\pi \ln z}r}{\lambda }\right) +\frac{1}{2}\pi -\text{Si}\left( \frac{4%
\sqrt{\pi \ln z}r}{\lambda }\right) \right) .  \label{3.21}
\end{eqnarray}%
where $j_{1}\left( x\right) $ is the spherical Bessel function of order $1$.

\subsection{Representation of degeneracy by scaling}

\begin{figure}[t]
\includegraphics[width=0.8\columnwidth]{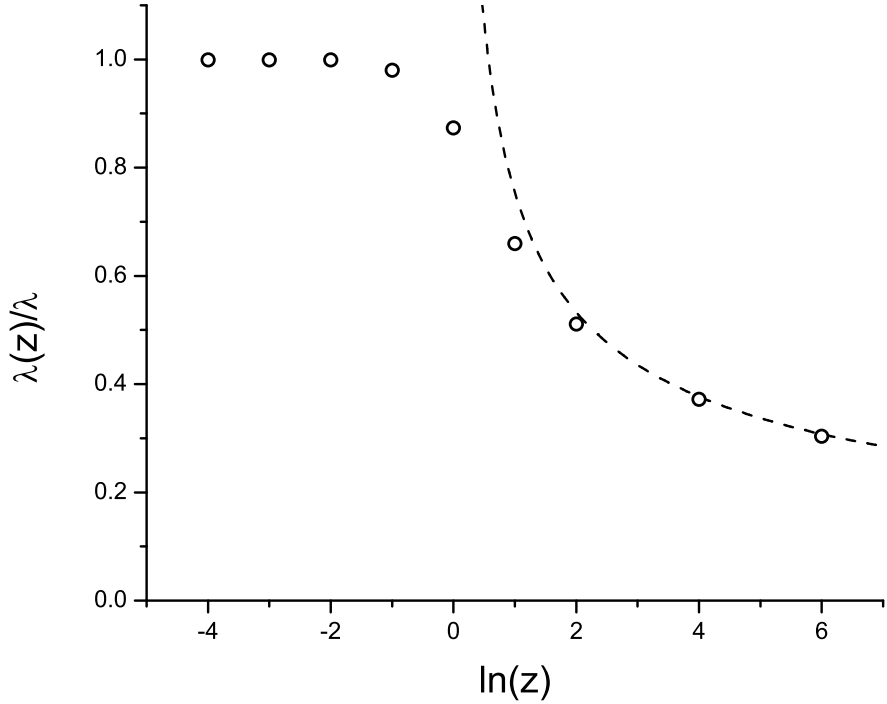} .
\caption{$\protect\lambda (z)/\protect\lambda $ as a function of $ln(z)$
(open circles). Also shown is the asymptotic limit proportional to $1/%
\protect\sqrt{ln(z)}$ (dashed line).}
\label{fig1}
\end{figure}

\begin{figure}[t]
\includegraphics[width=0.8\columnwidth]{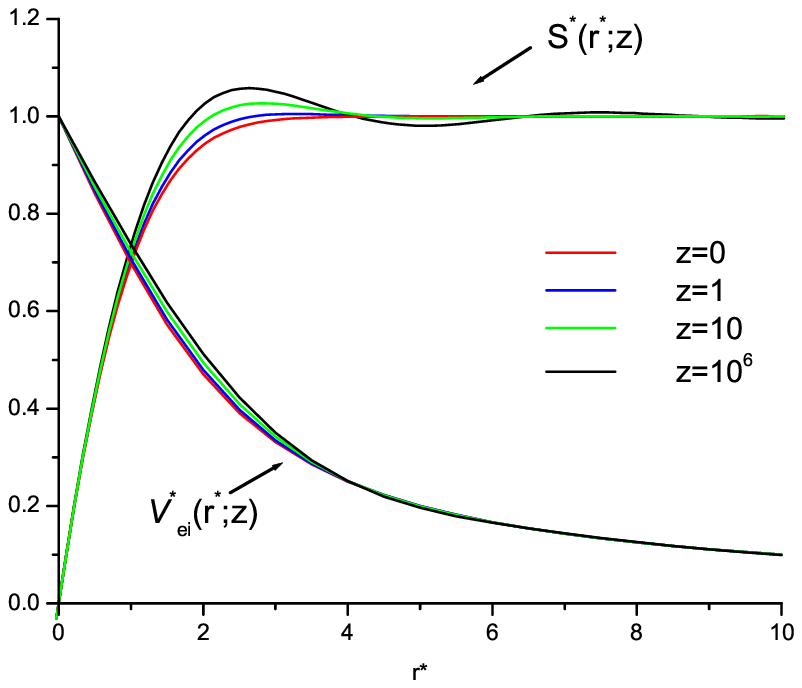} .
\caption{$S^{\ast}(r^{\ast};z)$ and
${\mathcal{V}_{ei}}^{\ast}(r^{\ast};z)$ as functions of $r^{\ast}$
for $z=0,1,10$ and $10^{6}$.} \label{fig2}
\end{figure}

It is interesting to note that the limiting forms (\ref{3.20}) and (\ref%
{3.21}) are both scaling functions, scaled by $\lambda $ in the first case
and by $\lambda /\sqrt{\ln z}$ in the second case. To explore the extent to
which effects of degeneracy can be described by scaling alone, consider the
degeneracy dependent wavelength $\lambda \left( z\right) $ defined in (\ref%
{3.19}). For small $z$ it approaches $\lambda $ while for large $z$ it is
proportional to $1/\sqrt{\ln z}$ as shown in Figure 1. Hence it is a
possible scaling length to interpolate between these limits. Accordingly,
define $S^{\ast }(r^{\ast },z)$ by
\begin{equation}
S^{\ast }(r^{\ast }\equiv \frac{r}{\lambda (z)},z)=S(\frac{r}{\lambda },z).
\label{3.22}
\end{equation}%
It follows from (\ref{3.19}) that this scaling assures that the initial
slopes of $S^{\ast }(r^{\ast },z)$ are the same for all $z$ . Figure 2 shows
the extent to which this scaling captures the effects of degeneracy for a
wide range of $z$. Also shown is the corresponding dimensionless quantum
potential $\mathcal{V}_{ei}{}^{\ast }(r^{\ast },z)\equiv {S}^{\ast }(r^{\ast
},z)/r^{\ast }$. For large and small $r^{\ast }$ the curves are the same,
although there are some differences for intermediate values of $r^{\ast }$.
This is due mainly to the oscillatory feature that develops for strong
degeneracy (related to Friedel oscillations). However, the quantitative
effect on the quantum potential in these scaled units is quite small.

This suggests the approximation for arbitrary degeneracy%
\begin{equation}
S^{\ast }(r^{\ast },z)\simeq S^{\ast }(r^{\ast },0),  \label{3.23}
\end{equation}%
or correspondingly, the approximate quantum potential%
\begin{equation}
\mathcal{V}_{ei}(r)\simeq V(r)S_{K}(\frac{r}{\lambda (z)}).  \label{3.24}
\end{equation}%
Here $S_{K}(r/\lambda (z))$ is the non-degenerate Kelbg form of (\ref{3.20}%
), but now with $\lambda $ replaced by $\lambda (z)$. Thus, approximation (%
\ref{3.24}) is a universal function for all degrees of degeneracy. The
change in length scale with degeneracy can be understood by noting that the
characteristic energy defining this scale is not $k_{B}T$ but rather the
average kinetic energy per particle which approaches the Fermi energy for
large $z$.

The above explicit results for $\mathcal{V}_{ei}(r)$ are limited to weak
coupling. In the case of an attractive ion at the origin there are important
bound state effects that are not included in this weak coupling form.
However, it has been shown \cite{filinov} that such strong coupling effects
can be included approximately by parameterizing the Kelbg form to fit the
exact value of $n(\mathbf{r=0},z=0,\beta )$. The possibility of extending
this to $z>0$ in (\ref{3.24}) will be explored elsewhere.

\section{Summary}

One of the simplest quantum systems exhibiting both diffraction and exchange
effects is a fixed impurity in an ideal Fermi gas of electrons. Here, a
classical system has been associated with that quantum system by the
introduction of two quantum potentials. The first is the well-known pair
interaction potential among the classical electrons to represent exchange,
while the second is a renormalization of the bare impurity-electron
interaction. The potentials are defined by the requirement that pair
correlations for the classical and quantum systems should be the same. The
classical pair potential is determined entirely by the ideal Fermi gas
correlation function and describes only exchange effects. The classical
electron-impurity potential differs from the bare potential of the quantum
system by both exchange and diffraction effects in a complex mixture of the
two. A simple representation at weak coupling is given by the familiar Kelbg
form for diffraction regularization, but modified by a degeneracy dependent
length scale.

Applications of classical molecular dynamics to real systems, such as a
hydrogen, require a classical representation with quantum potentials
representing both quantum effects and Coulomb interactions among all
particles. Current applications use quantum potentials for the electrons
that are the sum of an exchange potential $\mathcal{V}_{ee}$ as determined
here plus a regularized Coulomb potential of the Kelbg type for diffraction
effects. However, the analysis of the impurity problem here suggests that
exchange and diffraction are not likely to be additive. This is clear from
Eq. (\ref{2.8}) where all information about the quantum effects enters via $%
n_{ee}$ where all effects are mixed (e.g., in the random phase
approximation). It is only in the sense of perturbation in one or the other
that they become additive.

\section{Acknowledgements}

The research was supported by the NSF/DOE Partnership in Basic Plasma
Science and Engineering under the Department of Energy award
DE-FG02-07ER54946.


\bigskip

\begin{thebibliography}{99}
\bibitem{kelbg} G. Kelbg, Ann. Phys. (Leipzig) \textbf{12}, 219 (1963);
\textbf{13}, 354 (1963); \textbf{14}, 394 (1964).

\bibitem{PIMC} V.S. Filinov, M. Bonitz, W. Ebeling, and V.E. Fortov, Plasma
Phys. Cont. Fusion \textbf{43}, 743-759 (2001); V. Filinov et al. J. Phys.
A: Math. Gen. \textbf{36}, 6069 (2003).

\bibitem{feynman} R.P. Feynman and H. Kleinert, Phys. Rev. A \textbf{34},
5080 (1986); H. Kleinert, \emph{Path Integrals in Quantum Mechanics,
Statistics and Polymer Physics}, 2nd ed. (World Scientific, Singapore, 1995).

\bibitem{filinov} For recent reviews, see A.~Filinov, V. Golubnychiy,
M.~Bonitz, W.~Ebeling, and J.~Dufty, Phys. Rev. E \textbf{70}, 046411
(2004); A. Filinov, M. Bonitz, and W. Ebeling, J. Phys. A: Math. Gen.
\textbf{36}, 5957 (2003); W. Ebeling, A. Filinov, M. Bonitz, V. Filinov, and
T. Pohl, J. Phys. A: Math. Gen. \textbf{39}, 4309 (2006).

\bibitem{external} D.~Bohm, Phys. Rev. \textbf{85}, 166 and 180 (1986);
D.K.~Ferry and J. R.~Zhou, Phys. Rev. B \textbf{48}, 7944 (1993).

\bibitem{fromm} A. Fromm, M. Bonitz, and J. Dufty, Annals of Physics \textbf{%
323}, 3158 (2008).

\bibitem{Lado} F. Lado, J. Chem. Phys. \textbf{47}, 5369 (1967).

\bibitem{Hansen} J-P Hansen and I. MacDonald, \emph{Theory of Simple Liquids}%
, (Academic Press, San Diego, 1990).

\bibitem{Perrot} M.W. C. Dharma-wardana and F. Perrot, Phys. Rev. Lett.
\textbf{84}, 959 (2000); F. Perrot and M.W. C. Dharma-wardana, Phys. Rev. B
\textbf{62}, 16536 (2000); M. W. C. Dharma-wardana, Phys. Rev. Lett. \textbf{%
101} 035002 (2008).

\bibitem{Jones} C. Jones and M. Murillo, High Energy Density Physics \textbf{%
3}, 379 (2007).

\bibitem{fetter} A. L. Fetter and J. D. Walecka, \emph{Quantum Theory of
Many-Particle Systems} (McGraw-Hill, NY, 1963).

\bibitem{dufty} J. Dufty, S. Dutta, M. Bonitz, A. Filinov (unpublished).
\end{thebibliography}
\end{document}